\documentclass[prl,superscriptaddress,twocolumn,notitlepage,longbibliography]{revtex4-1}
\usepackage[unicode=true, bookmarks=true,bookmarksnumbered=false,bookmarksopen=false,
 breaklinks=false,pdfborder={0 0 1},backref=false,colorlinks=true]
 {hyperref}
\hypersetup{
linkcolor=magenta,urlcolor=blue,citecolor=blue,pdfstartview={FitH},urlcolor=blue}

\usepackage{amsfonts}
\usepackage{subfigure}
\usepackage{amsmath}
\usepackage{amssymb}
\usepackage{amsbsy} 
\usepackage{epsfig}
\usepackage{graphicx}
\usepackage{epstopdf}
\usepackage{bbm}
\usepackage{xcolor}

\def\be{\begin{equation}} \def\ee{\end{equation}}
\def\bea{\begin{eqnarray}} \def\eea{\end{eqnarray}}

\begin{document}

\title{Universal non-Hermitian transport in disordered systems}

\author{Bo Li}
\affiliation{MOE Key Laboratory for Nonequilibrium Synthesis and Modulation of Condensed Matter,\\
Shaanxi Province Key Laboratory of Quantum Information and Quantum Optoelectronic Devices,\\
School of Physics, Xi’an Jiaotong University, Xi’an 710049, China}
\affiliation{
Institute for Advanced Study, Tsinghua University, Beijing, 100084, China}

\author{Chuan Chen}
\affiliation{Lanzhou Center for Theoretical Physics, Key Laboratory of Quantum Theory and Applications of MoE,
Key Laboratory of Theoretical Physics of Gansu Province,
and School of Physical Science and Technology, Lanzhou University, Lanzhou, Gansu 730000, China}
\affiliation{
Institute for Advanced Study, Tsinghua University, Beijing, 100084, China}

\author{Zhong Wang}
\email{wangzhongemail@tsinghua.edu.cn}
\affiliation{
Institute for Advanced Study, Tsinghua University, Beijing, 100084, China}
	
\begin{abstract}

In disordered Hermitian systems, localization of energy eigenstates prohibits wave propagation. In non-Hermitian systems, however, wave propagation is possible even when the eigenstates of Hamiltonian are exponentially localized by disorders. We find in this regime that non-Hermitian wave propagation exhibits novel universal scaling behaviors without Hermitian counterpart. Furthermore, our theory demonstrates how the tail of imaginary-part density of states dictates wave propagation in the long-time limit. Specifically, for the three typical classes, namely the Gaussian, the uniform, and the linear imaginary-part density of states, we obtain logarithmically suppressed sub-ballistic transport, and two types of subdiffusion with exponents that depend only on spatial dimensions, respectively. Our work highlights the fundamental differences between Hermitian and non-Hermitian Anderson localization, and uncovers unique universality in non-Hermitian wave propagation.

\end{abstract}

\maketitle

The Anderson localization is one of the most prominent phenomena in random systems~\cite{Anderson1958}. In recent years, the rapid development of non-Hermitian physics has sparked growing interest in the interplay between non-Hermiticity and Anderson localization. This has led to the discovery of many intriguing phenomena unique to non-Hermitian systems, including non-Hermitian Anderson transition~\cite{Hatano1996NHmodel,Jiang2019NHSEquasiperiodic,Longhi2019TopoQuasicrystal,Longhi2019NHandersionTransition,Huang2020nhAnderson,Liu2020NHquasiperiodic,Zeng2020NHmobility,Kawabata2021NonunitaryScaling,Weidemann2022NHfloquetquasicrystal,Luo2021transfermatrixAT,Luo2021NHandersonTransition,Liu2021NHquasicrystal,Luo2022NHandersonTransition,Liu2024dissipationLocal,Wang2025nonBlochAT}, topological Anderson insulator~\cite{Zhang2020NHtopoAnderson,Tang2020TopoAnderson,Liu2021NHtopoAnderson2D,Lin2022experimentNHtopoAI}, many-body localization~\cite{Hamazaki2019nonHermitianMBL,Zhai2020MBLquasiperiodic,Suthar2022nhMBL,Wang2023skinMBL,Roccati2024nonHermitianMBLchaos}, Lifshitz tail states~\cite{Marchetti2001susyDOS,Silvestrov2001NHtailDOS,Longhi2025LifshitzTail}, universality of non-Hermitian random matrices~\cite{Hamazaki2020NHmatrixuniversality,Chen2025NHsigmaModel}, erratic non-Hermitian skin localization~\cite{Longhi2025erratic}, and anomalous dynamics~\cite{Balasubrahmaniyam2020Necklace,Weidemann2021NHtransport,Longhi2023LocLindblad,Yusipov2018AndersonJump, Tzortzakakis2021NHtransport, Leventis2022NHjump, Sahoo2022NHAndersonTransport,Tzortzakakis2020complexdisorder,Yusipov2017OpenSysLocalization}. 

The Anderson localization has twofold meanings. The spectral meaning is the localization of eigenstates, while the dynamic meaning is the absence of diffusion. It is generally perceived that the spectral localization always implies dynamical localization. For example, when the electron eigenstates at the Fermi level are Anderson-localized, electron transport is prohibited in a crystal, resulting insulating behavior at low temperature. In sharp contrast to this scenario, recent experiments in open or non-Hermitian systems have found that, despite complete localization of eigenstates, wave propagation remains possible~\cite{Weidemann2021NHtransport,Longhi2023LocLindblad} (see also related works Refs.~\cite{Yusipov2018AndersonJump, Tzortzakakis2021NHtransport, Leventis2022NHjump, Sahoo2022NHAndersonTransport,Tzortzakakis2020complexdisorder,Yusipov2017OpenSysLocalization}). In these systems, random gain and loss localize all the eigenstates, yet waves can propagate in a jumpy manner, dynamically evading the localization.

Experimental and numerical findings suggest that this unconventional non-Hermitian transport dramatically differs from typical ballistic or diffusive motions in terms of the relationship between spreading distance and time scale~\cite{Weidemann2021NHtransport}. It is also recognized that the complexity of eigen-energies is essential, indicating the intrinsically non-Hermitian nature of the phenomenon. However, it remains elusive whether the observed unusual dynamics obeys universal laws.  Here, we present a quantitative theory that yields universal transport dynamics, which not only explains the previous experimental findings but also predicts new scaling behaviors. The universality in this context manifests in the following aspects: (i) the space-time scaling of wave spreading is governed by the imaginary-part density of states, particularly its tail behavior; (ii) the scaling behavior depends solely on the dimensionality and the nature of the disorder, independent of other system-specific details; (iii) For weak disorders, the scaling approaches a universal form.
Our theory highlights the fundamental differences between Hermitian and non-Hermitian Anderson localization, and provides an efficient scheme for calculating the universal behaviors from the basic data of the system.   Our approach bears some similarity to Mott's variable-range hopping, though the mechanism is quite different \cite{Mott1969VRhopping}. Roughly speaking, the time plays the role of inverse temperature in Mott's picture.


\begin{figure}
    \centering
\includegraphics[width=1\linewidth]{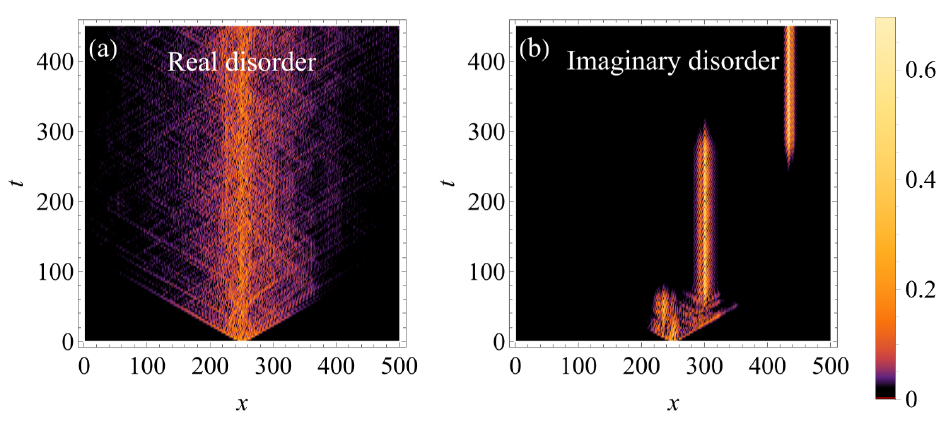}
\caption{The space-time evolution of a wave packet initially localized at the center of a chain is plotted under (a) real and (b) imaginary disorder, as described by Eq.~\eqref{eq:Hamiltonian} (For the real disorder scenario, the imaginary unit in front of the potential is omitted). Here, the disorder follows a uniform distribution $V_x\in [-W,W]$, and $t_0=2W=1$ is used in the plots.}
\label{fig:evolution}
\end{figure}

\textit{Dynamical scaling from an optimization}.
To be concrete, we consider a particle moving in a purely imaginary disorder potential
\begin{eqnarray}\label{eq:Hamiltonian}
H=\sum_{\langle\mathbf{x},\mathbf{x}^\prime\rangle} t_0 (|\mathbf{x}\rangle\langle \mathbf{x}^\prime|+h.c.)+ i\sum_{\mathbf x}V_{\mathbf x} |\mathbf x\rangle\langle \mathbf x|
\end{eqnarray}
where $\langle\mathbf{x},\mathbf{x}^\prime\rangle$ stands for the nearest sites on a square-lattice model, $t_0$ is real, and $V_{\mathbf x}$ is a real site-dependent random variable. This model naturally arises in various contexts; for example, it describes classical wave propagation in a lattice with random gain and loss~\cite{Weidemann2021NHtransport} and the dynamics of quantum particles in stochastic dissipative environments (see below). Note that our main results remain applicable if the random potential takes generic complex values, though we shall focus on the simplest cases with purely imaginary values. Numerical simulation of wave spreading exhibits a jumpy behavior [Fig.~\ref{fig:evolution}(b)], which is remarkably different from the real-valued-disorder case with wave confined in a localization length [Fig.~\ref{fig:evolution}(a)]

Similar to the standard Anderson model, the eigenstates of this Hamiltonian exhibit exponential localization~\cite{Basiri2014LightLocalization,Weidemann2021NHtransport}. However, the eigenvalues now take random complex values. We label the eigenvalues $E_{\mathbf x} (\in \mathbbm{C})$ and eigenstates $|\psi_{\mathbf x}\rangle$ by the localization center ${\mathbf x}$.  Suppose that a particle or wave packet is initially located at $\mathbf x=\boldsymbol{0}$. Its time evolution can be conveniently analyzed in the eigenbasis, i.e., $|\phi(t)\rangle=\sum_{\mathbf x} a_{\mathbf x}(t)|\psi_{\mathbf x}\rangle$, where the coefficient has modulus $|a_{\mathbf x}(t)|\sim e^{-|\mathbf x|/\xi+\lambda_{\mathbf x}t}$, $\xi$ being the eigenstate localization length and $\lambda_{\mathbf x}=\text{Im}(E_{\mathbf x})$. 
The wave intensity (or particle density) at a position $\mathbf{x}$ at time $t$ is  $P(\mathbf x,t)\sim  e^{2W(\mathbf x,t)}$, where $W(\mathbf x,t)=-|\mathbf x|/\xi+\lambda_{\mathbf x}t$ is a weight factor. At each moment, the average center
spreading distance $|\mathbf x_c|=\sum_{\mathbf x}|\mathbf x| P(\mathbf x,t)/\sum_{\mathbf x}P(\mathbf x,t)$ is dominated by the localization center of the eigenstate with maximum weight factor, 
and contribution from other eigenstates are exponentially suppressed. The weight factor measures the competition between the exponential tail suppression ($-|\mathbf x|/\xi$) and the temporal amplification ($\lambda_{\mathbf{x}}t$). As time grows, the most probable localization center moves to further positions,
spreading the wavepacket over space. Therefore, the average displacement is obtained by optimizing the weight factor at each moment~\cite{Mott1969VRhopping,Apsley1974Randhopping,Apsley1975Radomhopping,Brochard1977polymer,Zhang1986hopLocalization,BOUCHAUD1990Disorderdiffusion,Cates1988polymer,YCZhang1995Review}.

\begin{figure}
    \centering
\begin{tabular}{cc}
\includegraphics[width=0.5\linewidth]{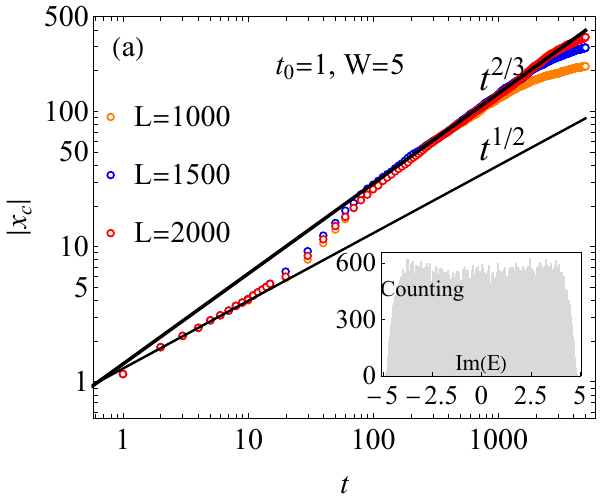}& 
\includegraphics[width=0.5\linewidth]{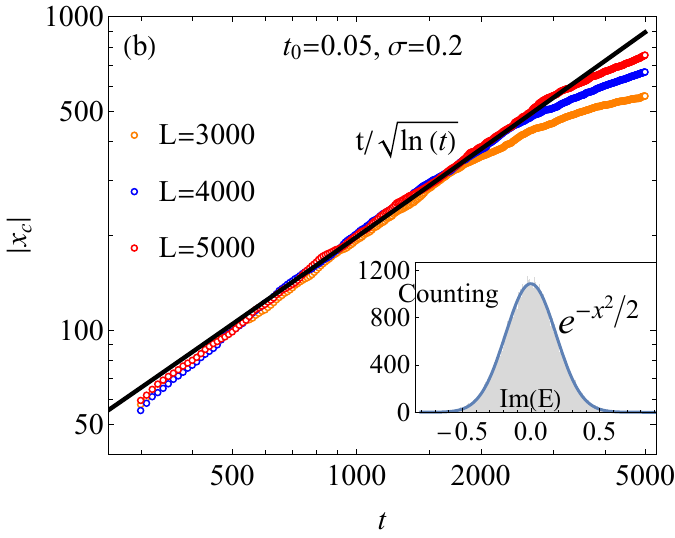}
\end{tabular}
\caption{Dynamical scaling in 1D lattice models with imaginary disorder. (a) For the uniform disorder, where $t_0=1$ and $W=5$, simulations are conducted with $500$ realizations of disorder. (b) Gaussian disorder with $t_0=0.05$ and $\sigma=0.2$ is considered, with 1500 realizations of disorder. The data for $\text{Im}(E)$ in both insets are collected from a system with $L=500$ and $200$ trials of disorder.}
\label{fig:unif_gau_scaling}
\end{figure}

To implement this optimization, we estimate the largest growing factor $\lambda^{\text{max}}_{|\mathbf x|}$ within a volume $\sim|\mathbf x|^d$. Since $\lambda_\mathbf{x}$ follows the same distribution across each site, $W(\mathbf x,t)$ is predominantly  optimized by $\lambda^{\text{max}}_{|\mathbf x|}$ within this volume. This is particularly relevant given the broad distribution of $\lambda_{\mathbf x}$ under strong randomness, leading to an exponential suppression of other contributions.  The probability of selecting a value greater than $\lambda^{\text{max}}_{|\mathbf x|}$ is  $P_{|\mathbf x|}=\int_{\lambda^{\text{max}}_{|\mathbf x|}}^\infty d\lambda \rho(\lambda)$, where $\rho(\lambda)$ is the imaginary-part density of states (ImDOS). Specifically, $\rho(\lambda)$ is defined as $\rho(\lambda)=\int_{-\infty}^{\infty} d\epsilon\rho(\epsilon,\lambda)$, where $\rho(\epsilon,\lambda)$  represents the density of states for the complex energy $\epsilon+i\lambda$ in the complex plane. In the considered region, there has to be at least one site taking the value $\lambda^{\text{max}}_{|\mathbf x|}$, implying $P_{|\mathbf x|} |\mathbf x|^d\sim 1$.  This yields~\cite{extreme}:


 

\begin{eqnarray}\label{eq:pxrelation}
 P_{|\mathbf x|}\sim |\mathbf x|^{-d},  
\end{eqnarray}
meaning that $P_{|\mathbf x|}$ should diminish as $|\mathbf x|$ grows.
This indicates that the long-time scaling behavior is dictated by the tail of ImDOS.  If ImDOS $\rho(\lambda)$ is specified, the relation between $\lambda^{\text{max}}_{|\mathbf x|}$ and $|\mathbf x|$ can be further identified with the help of Eq.~\eqref{eq:pxrelation}. For instance, if the ImDOS follows the Gaussian distribution $\rho(\lambda)\propto e^{-\lambda^2/(2\sigma^2)}$ with $\sigma$ being the standard deviation, by using the expansion of the error function we can obtain $\lambda^{\text{max}}_{|\mathbf x|}\sim (d\ln |\mathbf{x}|)^{1/2}$~\cite{Supp}. Substituting this relation into the weight factor and letting
\begin{eqnarray}\label{eq:optimization}
\frac{d W}{d|\mathbf{x}|}=-\frac{1}{\xi}+\frac{d\lambda_{|\mathbf{x}|}^{\text{max}}}{d|\mathbf{x}|}t =0,
\end{eqnarray} 
we arrive at the  space-time scaling of spreading distance~\cite{Supp}
\begin{eqnarray}\label{eq:GaussianScaling}
    |\mathbf x_c|\sim \frac{t}{ (\ln t)^{1/2}},
\end{eqnarray}
which implies a sub-ballistic motion. For other distributions, the resulting scaling behaviors are summarized in Table.~\ref{table_scaling}. Note that the scaling is determined by an integral of ImDOS with $\lambda^{\text{max}}_{|\mathbf x|}$ being the lower bound. As $\lambda^{\text{max}}_{|\mathbf x|}$ increases with spatial distance and time, the effect of ImDOS will gradually fade away, leaving its small tail to become dominant in late-stage dynamics. In this process, a crossover from one scaling to another may arise if the ImDOS does not take a universal functional form [e.g., the example in Fig.~\ref{fig:unif_gau_scaling} (a) discussed later].
\begin{table}[ht]
	\centering
	\begin{tabular}{|c| c| c|}
		\hline\hline 
	 ImDOS& $\rho(\lambda)$ & Dyanmical scaling \\ [0.5ex] 
		\hline 
	Gaussian& $e^{-\lambda^2/(2\sigma^2)}$ & $|\mathbf x_c|\sim t/(\ln t)^{1/2}$\\
 \hline
        Uniform& Constant  & $|\mathbf x_c|\sim t^{1/(d+1)}$\\
        \hline
        Linear& $ a-b\lambda$ & $|\mathbf x_c|\sim t^{2/(d+2)}$
		\\[1ex]
		\hline 
	\end{tabular}
	\caption{Dynamical scaling for different ImDOS.  For the linear ImDOS, $a,b$ are real parameters. The domain of $\lambda$ is given to ensure the normalization, not specified here. } \label{table_scaling}
\end{table}

To further clarify our approach, we compare it with Mott’s variable-range hopping (VRH)~\cite{Mott1969VRhopping}. In a strongly disordered system, electrons can hop between localized states a distance $R$ apart with the aid of phonons. At low temperatures, the hopping probability is estimated as $P\sim e^{-\frac{2R}{\xi}-\frac{\Delta}{k_BT}}$, where $\xi$ is the localization length and $\Delta$ is the energy difference between two states. The conductivity, governed by the most probable hopping process, results from an optimization between the spatial term $2R/\xi$, which favors short hops, and the energy term $\Delta/(k_BT)$, which benefits from larger hops to minimize energy differences. This optimization is achieved by approximating $\Delta$
as a function of $R$ and solving $\frac{\partial}{\partial R}[2R/\xi+\Delta(R)/(k_BT)]=0$~\cite{MottRelation}. Notably, despite the entirely different physical pictures, the optimization process in our approach exhibits similarities to VRH, where $1/(k_BT)$ in VRH plays a role analogous to $t$ in our dynamical scaling analysis.

\textit{1D examples with strong disorder}.
To validate our dynamic scaling predictions, we conducted numerical simulations for 1D cases. The center spreading distance $|x_c|=\sum_{\mathbf x}|x||\langle x|\phi (t)\rangle|^2/\langle\phi(t)|\phi(t)\rangle$ is numerically calculated
and then averaged over various disorder configurations. 
 In our initial exploration, we examined a uniform disorder potential with $V_{x}\in [-W,W]$.  For sufficiently strong disorder, $W\gg t_0$, dominating the imaginary part of spectrum, the ImDOS is approximately a uniform distribution [$\rho(\lambda)$= const.] except for the linearly 
decreasing tail [$\rho(\lambda)\sim a-b\lambda$] near the edge, as  depicted in the inset of Fig.~\ref{fig:unif_gau_scaling} (a). As a result, a $|x_c|\sim t^{1/(d+1)}|_{d=1}= t^{1/2}$ scaling appears in the early stage because $P_{|\mathbf x|}$ in Eq.~\eqref{eq:pxrelation} is mainly dictated by a uniform ImDOS. However, the late-stage scaling switches to $|x_c|\sim t^{2/(d+2)}|_{d=1}= t^{2/3}$ as a consequence of the linear tail of ImDOS.  This prediction aligns well with numerical simulations in Fig.~\ref{fig:unif_gau_scaling} (a), where size-effect-free scaling is evident from the overlap between the curves for different system sizes. The deviation from the thermodynamic-limit trajectory is observed to be delayed for larger systems compared to smaller ones.  Here, the linear ImDOS tail can be understood by the rare events of neighboring sites occupied by potentials with similar values~\cite{Supp}. Moreover, our theory and simulations are fully consistent with the optical experiment~\cite{Weidemann2021NHtransport}.
Another significant case involves Gaussian-type ImDOS, generated by strong Gaussian imaginary disorder with $\sigma\gg t_0$. As shown in Fig.~\ref{fig:unif_gau_scaling} (b), the numerical data for the  larger-size system overlay on top of the smaller one until size effect entering the dynamics. The theoretical prediction in Eq.~\eqref{eq:GaussianScaling} matches perfectly with the overlapped part at a large time, owing to that Eq.~\eqref{eq:GaussianScaling} is justified for large $\lambda$.

\begin{figure}
    \centering
    \begin{tabular}{cc}
\includegraphics[width=0.5\linewidth]{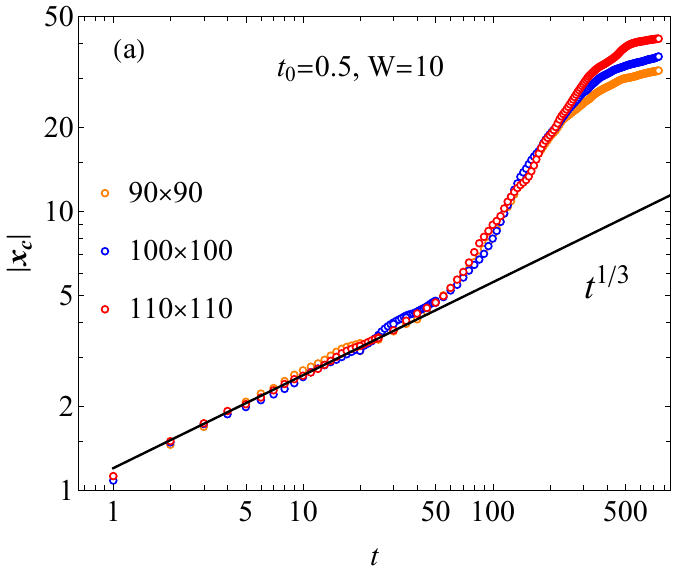}&
\includegraphics[width=0.5\linewidth]{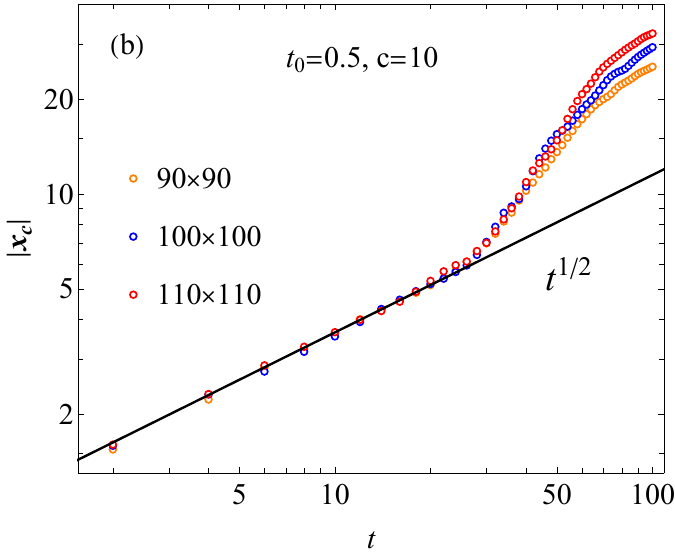} 
    \end{tabular}
    \caption{Dynamical scaling for 2D square-lattice models with system size given by the plot legends, where $500$ realizations of disorder are performed. (a) A uniform disorder $V_{\mathbf x}\in[-W,W]$ is considered. (b) The disorder follows a right triangle distribution, where $c$ is the length of the base line, see the text. }
    \label{fig:2D_scaling}
\end{figure}

\textit{Numerical results for 2D}.
In Table~\ref{table_scaling}, the uniform or linear ImDOS result in a dimension-dependent scaling, providing a key feature for verifying our predictions. Similar to the 1D case, a strong uniform disorder in higher dimension is expected to give an ImDOS with a linear tail attached to a uniform bulk part. In Fig.~\ref{fig:2D_scaling} (a), we performed simulations for the 2D case, where $|\mathbf x_c|\sim t^{1/(d+1)}|_{d=2}=t^{1/3}$ is indeed observed in the early dynamical stage. However, constraints on simulation size prevent the observation of late-stage dynamics influenced by the linear ImDOS tail. To address dimension sensitivity stemming from a linear ImDOS, simulations are conducted under the influence of strong imaginary disorder adhering to a (right angled) triangular distribution,  i.e., the probability density follows $f(V_{\mathbf x})=2(1-V_{\mathbf x}/c)/c$ with $V_\mathbf x\in[0,c]$ and $t_0\ll c$. Notably, the predominant linear ImDOS profile facilitates rapid manifestation prior to the onset of size effects, thus leveraging the limited size in 2D simulations effectively. In Fig.~\ref{fig:2D_scaling} (b), the expected scaling $|\mathbf{x}_c|\sim t^{2/(d+2)}|_{d=2}=t^{1/2}$ is evident during the early stage, while long-term dynamics are influenced by both size effects and the ImDOS tail, which is not of interest here.

\textit{Universal scaling under weak disorder}. In the preceding examples, we focused on the strong-disorder regime, where the imaginary part of the density of states (ImDOS) is predominantly shaped by the statistical properties of the disorder.  As the disorder strength decreases, however, a universal Gaussian-type form of the ImDOS gradually emerges. In the weak-disorder limit, the localization length $\xi$ spans many unit cells ($\xi \gg 1$), and within this length scale, particle spreading exhibits ballistic scaling, characterized by $|\mathbf{x}| \sim t$. Nonetheless,  the earlier analysis remains applicable after coarse-graining and rescaling the length scale by $\xi$: the wave packet remains strongly localized in the rescaled system.
Despite the coarse-grained unit cell is occupied by many eigenstates, the long-time ($t\gg 1$) probability of the wave packet locating at the new cell positioning at $\mathbf{X}$ is mostly governed by the maximum growing factor $\lambda_{\mathbf X}$ therein, represented as $P(\mathbf X,t)\sim e^{-2|\mathbf X|/\xi+2\lambda_{\mathbf X}t}$, where $\lambda_{\mathbf X}$ follows the same distribution as the original $\lambda_{\mathbf x}$. Moreover, for sufficiently weak disorder, i.e., $\xi\gg 1$, the central limit theorem implies that $\lambda_{\mathbf x}$, which is  approximated as an average of $V_{\mathbf x}$ within the volume of size $\xi^d$, follows a Gaussian distribution~\cite{Supp}. Consequently, the scaling behavior converges to the universal form given in Eq.~\eqref{eq:GaussianScaling} in the weak-disorder, thermodynamic limit.  Nevertheless, its numerical validation necessitates exceptionally large system sizes, which will be left for future study.

The emerging Gaussian distribution is contingent upon condition $\xi\gg 1$, which gradually breaks down as disorder strength increases until the system approaches the regime of strong disorder, i.e. $\xi\lesssim 1$. In the case of intermediate disorder characterized by $W,\sigma\sim t_0$, the ImDOS cannot be sharply identified. Consequently, the associated scaling behavior undergoes an unknown crossover bridging the two extreme cases~\cite{Supp}.

\textit{Renormalization group analysis}. As explained above, the scaling behavior exhibits certain sensitivity to disorder strength. In the regime of weak disorders, our approach necessitates a rescaled interpretation. In this respect, it is helpful to view our problem from the renormalization group perspective. This not only validates our rescaled treatment for weak disorder cases but also suggests new scaling behaviors for the delocalized phase that is inaccessible to the previous approach. In long-wavelength limit,  the Hamiltonian Eq.~\eqref{eq:Hamiltonian} is represented as a Schr\"odinger equation $i\partial_t\psi(\mathbf x,t)=[-D\nabla^2+iV(\mathbf x)]\psi(\mathbf{x},t)$, where the potential obeys $\langle V(\mathbf x)V(\mathbf x^\prime)\rangle=v\delta^d(\mathbf x-\mathbf x^\prime)$ with $v>0$ representing the disorder strength. Performing the transformation $\psi(\mathbf x,t)=\exp(\Phi(\mathbf x,t))$, the Schr\"odinger equation is converted to a nonlinear equation \begin{eqnarray}\label{eq:nonlinear}
\partial_t\Phi=iD\nabla^2\Phi+i\eta(\mathbf{\nabla}\Phi)^2+V(\mathbf{x}),
\end{eqnarray} 
where $D=\eta=t_0$.
This form allows us to implement a dynamical renormalization group analysis~\cite{Ma1975CriticalFM,Forster1977randomfluid}. Following the standard procedure~\cite{Medina1989BurgereqRG}, we obtain a flow equation~\cite{Supp} 
\begin{eqnarray}
\frac{dg}{dl}=g[4-d+\frac{2(12-5d)}{d}K_dg],
\end{eqnarray}
where $g=v\eta^2/D^4$, $K_d=S_d/(2\pi)^d$ with $S_d$ being the solid angle in $d$ dimension, and the rescaling $\mathbf x\rightarrow e^l\mathbf x$ is performed.
The flow equation yields two fixed points, $g_1^\ast=0$, $g^\ast_2=d(d-4)/[2(12-5d)K_d]$. Notably, $g_1^\ast$ is unstable for $d<4$, while $g^\ast_2$ is physically  untenable  for $d< 12/5$  due to its negative value, conflicting with the definition. Thus, in 1D and 2D, the system does not transit to new phases upon rescaling, validating our analysis for weak disorders.  However, for $d=3$, $g^\ast_2=1/(2K_d)$ is a genuine fixed point, signifying a delocalization transition.  In the delocalized phase with weak disorder, the system's long-time behavior is governed by this fixed point, meaning the space-time relation remains invariant under rescaling. Consequently, the system exhibits dynamical scaling $|\mathbf x|\sim t^{3/5}$~\cite{Supp}. This result is independent of the disorder statistics and beyond the approach outlined in Eqs.~\eqref{eq:pxrelation} and~\eqref{eq:optimization} which focuses on the eigenstate-localized regime.


\textit{Liouvillian dynamics.} Although the preceding analysis focuses on systems with non-Hermitian Hamiltonians, it is noteworthy that analogous dynamics can arise within the framework of the Liouvillian superoperator.
We consider an open system described by the following Lindblad master equation of density matrix $\hat\rho$
\begin{eqnarray}\label{eq:master}
\frac{d\hat \rho}{d t}=-i[\hat H_0,\hat \rho]+\sum_{\mathbf x}\gamma_{\mathbf x}\mathcal{D}[\hat a_{\mathbf x}]\hat \rho  
\end{eqnarray}
where $\hat H_0=\sum_{\mathbf x,\mathbf x^\prime}h_{\mathbf x,\mathbf{x}^\prime}\hat a^\dagger_{\mathbf x}\hat a_{\mathbf x^\prime}$ is a Hermitian Hamiltonian, with $\hat a^\dagger_\mathbf{x} (\hat a_{\mathbf x})$ being the creation (annihilation) operator at lattice site $\mathbf x$, and $\mathcal D[\hat a_{\mathbf x}]\hat\rho=\hat a_{\mathbf x}\hat\rho\hat a_{\mathbf x}^\dagger-\frac{1}{2}(\hat a_{\mathbf x}^\dagger\hat a_{\mathbf x}\hat\rho+\hat\rho\hat a^\dagger_{\mathbf x}\hat a_{\mathbf x})$.  On the right hand side of Eq.~\eqref{eq:master}, the first term describes a unitary evolution given by $\hat H_0$, and the second term accounts local losses with site-dependent random rate $\gamma_{\mathbf x}$. Using Eq.~\eqref{eq:master}, we can show that two-point correlation function $G_{\mathbf x_1,\mathbf x_2}=\text{Tr}[\hat\rho\hat a_{\mathbf x_1}^\dagger \hat a_{\mathbf x_2}]$ satisfies the following matrix equation~\cite{Supp} 
\begin{eqnarray}
 \frac{dG}{dt}=i( H G-G H^\dagger)
\end{eqnarray} 
where $G=\sum_{\mathbf x,\mathbf x^\prime}G_{\mathbf x,\mathbf x^\prime}|\mathbf x\rangle\langle \mathbf x^\prime|$, and $H=h^T+iV$ describes a Hermitian system $h^T$ subject to a random imaginary disorder $iV$ with $V_{\mathbf x,\mathbf x^\prime}=\gamma_{\mathbf x}\delta_{\mathbf x,\mathbf x^\prime}$. The equation is solved by $G(t)=e^{iHt}G(0)e^{-iH^\dagger t}$. Considering $N$ bosonic particles (e.g., photons in a cavity) initially locate at $\mathbf x=\boldsymbol{0}$, i.e., $G_{\mathbf x,\mathbf x^\prime}(0)=N\delta_{\mathbf{x},\mathbf x^\prime}\delta_{\mathbf x,\boldsymbol{0}}$, the correlation matrix at time $t$ evolves to $G(t)=N|\phi(t)\rangle\langle\phi(t)|$
where $|\phi(t)\rangle=e^{iHt}| \boldsymbol{0}\rangle$ acts as a ``quantum state" dictated by the ``Hamiltonian" $H$. The probability distribution of surviving particles  is given by $P(\mathbf x,t)=G_{\mathbf x,\mathbf x}(t)/\sum_{\mathbf x}G_{\mathbf x,\mathbf x}(t)=|\langle\mathbf x|\phi(t)\rangle|^2/\langle\phi(t)|\phi(t)\rangle$. This coincides exactly with the particle evolution before, indicating the same dynamical scaling behavior in such an open system. Furthermore, similar dynamic behaviors can emerge from purely dissipative systems ($\hat H_0=0$) by considering suitable dissipators involving  adjacent sites~\cite{Supp,Longhi2023LocLindblad}.

\textit{Conclusion}. In systems featuring complex disorders, the non-unitary time evolution facilitates a particle's jumpy motion even amidst eigenstate localization, in sharp contrast to the conventional Anderson localization. Employing an intuitive optimization approach, we unveiled the universal dynamical scaling in this unconventional non-Hermitian transport, distinctly different from typical diffusive or ballistic transport. More interestingly, we find a close connection between the scaling behavior and the imaginary-part density of states (ImDOS), especially the long-time behavior is dictated by the tail of ImDOS. Our findings underscore the fundamental distinction between non-Hermitian and Hermitian localization, particularly in terms of dynamics. Numerous open questions persist, including the theoretical prediction and experimental observation of dynamic behaviors arising from different types of disorders, particularly in the regime with comparable kinetic and potential terms. It is also interesting to explore the combined dynamical consequence of Anderson localization and non-Hermitian skin effect~\cite{Yao2018NHSE,yao2018chern,kunst2018biorthogonalBBC,Xiao2020NHbulkboundary,Helbig2020generalBBC,Martinez2018NHlocal,Lee2019NHSEanatomy,Ashida2021Non,Bergholtz2021RMP,Gohsrich2024Perspective,Jiang2019NHSEquasiperiodic,Weidemann2022NHfloquetquasicrystal,Zeng2020NHAubry,Wang2025boundarylocalization}. Additionally, the interplay between interactions and non-Hermitian disorders poses an intriguing avenue for exploration. On the experimental front, photonic systems provide a promising platform, as demonstrated by relevant experiments~\cite{Weidemann2021NHtransport}. Furthermore, open systems featuring local random losses, such as dissipative cavity arrays, offer additional opportunities to validate our theory.


\textit{Acknowledgements}. This work is supported by National Key R\&D Program of China (No. 2023YFA1406702) and NSFC under Grant No. 12125405. B. L. is partially supported by NSFC under Grant No. 12404185. C.C. acknowledges support from NSFC under Grant No. 12347107.

\bibliographystyle{apsrev4-1-title}
\bibliography{disorder}

\begin{thebibliography}{69}
\bibitem [{Mot()}]{MottRelation}%
  \BibitemOpen
  \href@noop {} {}\bibinfo {note} {Here $\Delta(R)$ is estimated as the typical
  level spacing within a volume $R^3$. We require that at least one state falls
  within the energy range $\Delta (R)$ at the Fermi energy, which translates to
  the condition $\Delta(R)R^3\nu\sim 1$, where $\nu$ is the density of state at
  the Fermi energy. This results in the expression $\Delta(R)\sim
  \frac{1}{R^3\nu}$. Substituting this into the optimization equation, one can
  determine the optimized hopping probability and subsequently derive Mott’s
  relation for conductivity: $\sigma\sim e^{-(T_0/T)^{1/4}}$.}\BibitemShut
  {Stop}%
\end{thebibliography}

\end{document}


\title{Supplemental Material for ``Universal non-Hermitian transport in disordered systems"}

\author{Bo Li}
\affiliation{MOE Key Laboratory for Nonequilibrium Synthesis and Modulation of Condensed Matter,\\
Shaanxi Province Key Laboratory of Quantum Information and Quantum Optoelectronic Devices,\\
School of Physics, Xi’an Jiaotong University, Xi’an 710049, China}
\affiliation{
Institute for Advanced Study, Tsinghua University, Beijing, 100084, China}

\author{Chuan Chen}
\affiliation{Lanzhou Center for Theoretical Physics, Key Laboratory of Quantum Theory and Applications of MoE,
Key Laboratory of Theoretical Physics of Gansu Province,
and School of Physical Science and Technology, Lanzhou University, Lanzhou, Gansu 730000, China}
\affiliation{
Institute for Advanced Study, Tsinghua University, Beijing, 100084, China}

\author{Zhong Wang}
\email{wangzhongemail@tsinghua.edu.cn}
\affiliation{
Institute for Advanced Study, Tsinghua University, Beijing, 100084, China}

\maketitle

\onecolumngrid

\title{Supplemental Material for ``Universal dynamical scaling in Non-Hermitian Anderson transport"}

\author{Bo Li}
\affiliation{
Institute for Advanced Study, Tsinghua University, Beijing, 100084, China}

\author{Chuan Chen}
\affiliation{
Institute for Advanced Study, Tsinghua University, Beijing, 100084, China}

\author{Zhong Wang}
\email{wangzhongemail@tsinghua.edu.cn}
\affiliation{
Institute for Advanced Study, Tsinghua University, Beijing, 100084, China}

\maketitle

\onecolumngrid

\tableofcontents

\section{Dynamical scaling from a Gaussian $\text{ImDOS}$}
Here, we elaborate on the replacement-time scaling that arises from a Gaussian ImDOS $\rho(\lambda)\sim e^{-\lambda^2/(2\sigma^2)}$. Eq.(2) in the main text is specified as below
\begin{eqnarray}
P_{|\mathbf x|}=\int_{\lambda_{|\mathbf x|}^{\text{max}}}^{\infty}e^{-\lambda^2}d\lambda\sim |\mathbf x|^{-d}.
\end{eqnarray}
When a large distance is considered, $|\mathbf{x}|\gg 1$, $\lambda_{|\mathbf x|}^{\text{max}}$ should be much greater than unity ($t_0=1$ is assumed), so that the integral can be estimated by the leading order of expansion around $\infty$, 
\begin{eqnarray}
    e^{-(\lambda_{|\mathbf x|}^{\text{max}})^2}\frac{1}{\lambda_{|\mathbf x|}^{\text{max}}}\sim |\mathbf x|^{-d}.
\end{eqnarray}
This leads to
\begin{eqnarray}
    \lambda_{|\mathbf x|}^{\text{max}}\sim \Big(d\ln|\mathbf x|-\ln\lambda_{|\mathbf x|}^{\text{max}}+\cdots\Big)^{1/2},
\end{eqnarray}
which generates the leading-order result
\begin{eqnarray}\label{eq:gaussian_x_e}
    \lambda_{|\mathbf x|}^{\text{max}}\sim (d\ln|\mathbf x|)^{1/2}.
\end{eqnarray}
Then, substituting this expression into the optimization equation $\partial W(\mathbf{x},t)/\partial |\mathbf{x}|=0$ gives
\begin{eqnarray}
    |\mathbf x|(\ln |\mathbf{x}|)^{1/2}=\frac{\xi d^{1/2}}{2} t,
\end{eqnarray}
which further yields
\begin{eqnarray}
    \ln |\mathbf{x}|=\ln t-\frac{1}{2}\ln\ln |\mathbf x|+\text{const.}=\ln[t/(\ln t)^{1/2}]+\cdots
\end{eqnarray}
where ``$\cdots$" represents higher-order contributions and a scaling-irrelevant constant. In summary, the dominant scaling reads
\begin{eqnarray}
    |\mathbf x_c|\sim t/(\ln t)^{1/2}.
\end{eqnarray}
Note that the derivation above is only valid for large-distance scaling. Therefore, we anticipate that this relation agrees well with numerical results for $|\mathbf x|\gg 1$.

\section{The imaginary-part density of state ($\text{ImDOS}$)}

In this section, we study the imaginary-part density of state (ImDOS) in the strong and weak disorder limit. In the strong limit for a bounded disorder, the ImDOS is dictated by the disorder distribution, attached with some tails.  Here, we focus on the uncorrelated uniform disorder $V\in [-W,W]$ to reveal the origin of the linear tail. Corresponding conclusions for a Gaussian disorder will be commented on accordingly. In the weak limit, the emergence of Gaussian-type ImDOS is attributed to the central limit theorem and is independent of the specific form of disorder. However, when the disorder is comparable to the kinetic term, no simple conclusion can be obtained. Reliable results in this regime might require consulting some advanced random matrix theories, which would be left for future work.

\subsection{The strong-disorder limit ($W\gg t_0$)---the linear tail in a bounded disorder} 

In the case of strong disorder ($W\gg t_0$), the hopping term can be regarded as a perturbation to the disorder. In the zeroth order, the local potential value is the eigenvalue. When hopping terms are added, the corrections start from its second order because an off-diagonal perturbation can not contribute in the leading order, given that the zeroth-order eigenstates are strictly localized. The effect of hopping terms is mixing neighboring (imaginary) potentials to reach some values in between. A naive inference is that, if the magnitude of potential is bounded, the possibility of reaching maximum or minimum imaginary value is less than that of moderate values. This is because there are less choices to involve a larger potential than the target value. An extreme case is $V_i=W$, hopping terms can only reduce the magnitude of the eigenvalue, thus the possibility of arriving at $\varepsilon=iW$ is zero. 

Depending on the magnitude of fluctuations of neighboring potentials, the imaginary part of eigenvalues can be understood under the framework of either normal or degenerate perturbation theory. For simplicity, we take the 1D case as an example; results for higher dimensions can be understood in the same way.  For an eigenstate with $\varepsilon_i^{(0)}=i V_i$ (with $V_i\in[-W,W]$), if the local potential of its neighboring sites are not too close, i.e., $|V_{i\pm 1}-V_i|\gg t_0$,  the corrected eigenvalue reads (for 1D)
\begin{eqnarray}\label{eq:perturbation1}
    \varepsilon_i=iV_i-\frac{it_0^2}{V_i-V_{i+1}}-\frac{it_0^2}{V_i-V_{i-1}}+\cdots.
\end{eqnarray}
With sample manipulation under the assumption $|V_{i\pm 1}-V_i|\gg t_0$, it is easy to show the resultant $\varepsilon_i$ is bounded by the maximum and minimum value of $\{V_{i-1},V_i,V_{i+1}\}$. In particular, for $\varepsilon^{(0)}_i=iW$, its imaginary part  can be only reduced by corrections from hopping, and, to $O(t_0^2)$, the maximum possible resultant value is 
\begin{eqnarray}
    \max \text{Im}\varepsilon_i=W-\frac{t_0^2}{W}.
\end{eqnarray}
For $\lambda\in [W-t_0^2/W,W]$, the growing factor cannot be obtained by the perturbation theory above. Because $t_0^2/W\ll t_0\ll |V_i-V_{i\pm 1}|$, if the above theory can yield the wanted $\lambda$, there must be one or two sites taking potential value far from this range (compare to the width $t_0^2/W$ and $t_0$) involved to mix with a potential in this range (potentials within this range will not mix by assumption); as a result, the obtained growing factor must be out of this range, i.e., $\lambda=V_i-t_0^2/(V_i-V_{i+1})-t_0^2/(V_i-V_{i-1})<W-t_0^2/W$ with $V_i\in [W-t_0^2/W,W]$ and $V_i>V_{i\pm 1}$ to satisfy the precondition of perturbation formula above. 

We are interested in the tail density state, i.e, $\rho(\Delta)$ with $\Delta=W-\lambda$ being very small, say less than $t_0$. In particular, when $\Delta<t_0^2/W$, the target value can not be reached by Eq.~\eqref{eq:perturbation1}; the eigenvalue in this range can only be understood under the framework of the degenerate perturbation theory, namely, consider the case $|V_i-V_{i+1}|\lesssim t_0$.  This corresponds to the rare events that a small region is fully occupied by potential with similar magnitude~\cite{Silvestrov2001NHtailDOS}. In this region, particles are almost free from disorder; thus, hopping terms produce the real part of the eigenvalue, and the imaginary part is the average of the potential in this region. Assuming that the local fluctuation of potential in this region is $\delta<t_0$, and the region size is $R^d$, then the possibility of such a rare event is 
\begin{eqnarray}
    P\sim (\frac{\delta}{2W})^{R^d},
\end{eqnarray}
which decreases exponentially as $R^d$ increases. To capture the leading feature of $\rho(\Delta)$, we consider the case with $R=2$ and $d=1$ for simplicity, say potential $V_1, V_2$ on neighboring sites, then degenerate perturbation in the subspace yields
\begin{eqnarray}
    \varepsilon=i\frac{V_1+V_2}{2}\pm\sqrt{t_0^2-\frac{(V_1-V_2)^2}{4}}.
\end{eqnarray}
It is obvious that for a given $\Delta$ there is only one freedom of choosing local potential within these two sites; suppose the larger one is counted (the larger one is greater than the target value $\lambda=W-\Delta$), then the following relation holds  
\begin{eqnarray}
    \rho(\Delta)\sim \int_{W-\Delta}^W \frac{d\lambda}{2W}\propto\Delta.
\end{eqnarray}
If a larger $R$ is considered, more freedoms appear, and higher orders of $\Delta$ will enter the density of state. These contributions will be exponentially suppressed, and higher-order terms are overwhelmed by the linear term.\\

For $\Delta$ small but greater than $t_0^2/W$, the main contribution still comes from Eq.~\eqref{eq:perturbation1} (the degenerate-perturbation contribution is exponentially suppressed), so that there are two freedoms (in 1D) of choosing the value of involved local potential and the resultant probability is the joint probability of these two:
\begin{eqnarray}
    \rho(\lambda)&&\sim  \int_{\lambda}^W d\lambda_1 P(V_{\max}=\lambda_1)
    \int d\lambda_2 P(\text{one of the rest value}=\lambda_2|V_{\max}=\lambda_1)\nonumber\\
    &&=\int_{\lambda}^W \frac{d\lambda_1}{2W}\int_{-W}^{B(\lambda_1,\lambda)}\frac{d\lambda_2}{2W}\nonumber\\
    &&=\frac{W(W-\lambda)}{(2W)^2}+\frac{1}{(2W)^2}\int d\lambda_1B(\lambda_1,\lambda)\nonumber\\
    &&\sim \Delta+\mathcal O(\Delta^2)
\end{eqnarray}
where $\lambda=W-\Delta\sim W$. Here, it is not hard to show that the lower bound of $\lambda_2$ is $-W$, and the upper bound is a function of $\lambda$ and $\lambda_1$. The coefficient in front of $\Delta$ in the last line is not explicitly given because the second term in the second line from the bottom could also produce a linear term. In particular, the higher-order terms are highly suppressed near the band edge; thus, a linear behavior for $\rho(\lambda)$ is seen near the edge if the imaginary part of the spectrum is bounded. 

\subsection{The weak-disorder limit ($W\ll t_0$)---emergent Gaussian distribution}

In a weak disorder, the localization length is large, so that the eigenstates cover many sites with number $\sim \xi^d$, where $\xi$ is generally a function of eigenvalue $\varepsilon$ and mainly relies on $\text{Re}(\varepsilon)$ for weak non-Hermitian disorder. This picture is especially valid for these states in the middle of the real part of the band, i.e., $\text{Re}(\varepsilon)\approx 0$. The corresponding imaginary part of energy is asymptotically given by the average of the local potential in these sites, which follows a Gaussian distribution, according to the central limit theorem. This can be seen explicitly from a perturbative perspective. In 1D, near the band center, the unperturbed wave function is $\psi^{(0)}_j\sim e^{ik j}$ with $k\simeq\pm\pi/2$, and the imaginary disorder makes the eigenstate become a linear combination of states in a narrow region near $k$, i.e.,
\begin{eqnarray}
    \psi^R\approx\sum_{q\in [k-\epsilon,k+\epsilon]} c_q\psi_{q}^{(0)},\qquad \epsilon\sim W/t_0\ll 1.
\end{eqnarray}
Here, the narrow region works much better around the band center than the band edge. From perturbation theory, the linear combination should include all bands within a range $W/t_0$. For $k=\pm\pi/2+\epsilon$, $\delta E(k)= E(\pm\pi/2+\epsilon)-E(\pm\pi/2)\sim t_0\epsilon\sim W$, so $\epsilon\sim W/t_0$; however, around the band edge ($k=0$ or $\pi$), $\delta E=E(\epsilon)-E(0)\sim t_0\epsilon^2\sim W$, which yields $\epsilon\sim (W/t_0)^{1/2}$. For $W/t_0\ll 1$, the state near the band edge needs to involve many more states. Turning to real space, this means the states near the band center have a longer localization length, while those near the edge usually are sharper because the reciprocal space components for a narrow peak should have a broader distribution, e.g., the Dirac-delta function is a superposition of plane waves with all momentum components.

\begin{figure}
    \centering
   \begin{tabular}{cc}  \includegraphics[width=1\linewidth]{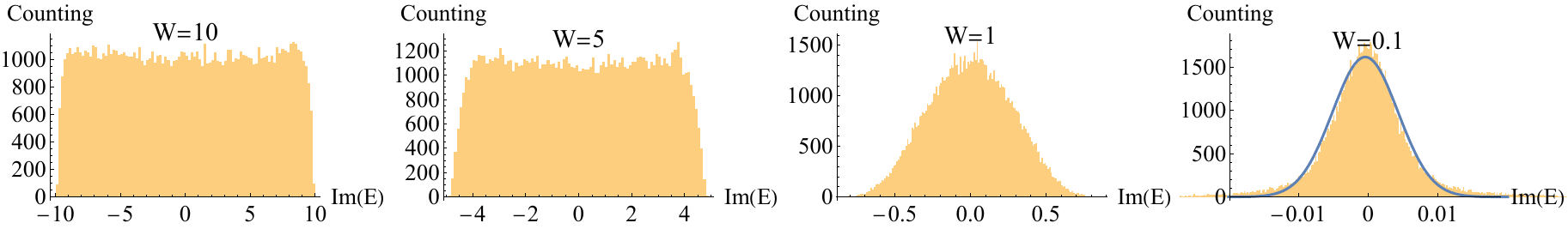}  \\ \includegraphics[width=1\linewidth]{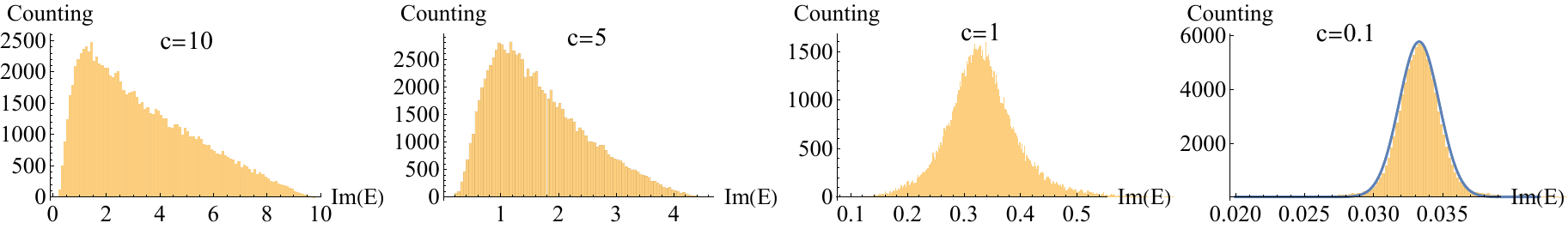} 
   \end{tabular}
    \caption{Emergence of Gaussian-type ImDOS with decreasing disorder strength in 1D. The histograms show the distribution of the imaginary parts of the eigenvalues. Top panels: results for uniformly distributed disorder, $V_x\in[-W,W]$. Bottom panels: results for triangularly distributed disorder with probability density $f(V_x)=\frac{2}{c}(1-V_x/c)$, where $V_x\in [0, c]$. In the rightmost panels, solid lines represent Gaussian fits. All histograms are generated with $t_0=1$, system size $L=500$, and averaged over 200 disorder realizations.}
    \label{fig:GuassianEmergence}
\end{figure}

The property $H^\dagger=H^\ast$ enforces $\langle\psi^L|=(|\psi^R\rangle)^T$, thus the normalization relation is given by
\begin{eqnarray}
\langle\psi^L|\psi^R\rangle=\sum_j\sum_{q,q^\prime}c_q c_{q^\prime}\psi_{q,j}^{(0)}\psi_{q^\prime,j}^{(0)}=\sum_q c_q^2=1.
\end{eqnarray}
Taking disorder into account, the eigenvalues is 
\begin{eqnarray}
\varepsilon=\varepsilon_1+\varepsilon_2=\langle\psi^L|H_0|\psi^R\rangle+i \langle\psi^L|V|\psi^R\rangle.
\end{eqnarray}
Here,  
\begin{eqnarray}
\varepsilon_1=\sum_q c_q^2\varepsilon^{(0)}_q\simeq \varepsilon_k^{(0)}\sum_qc_q^2=\varepsilon_k^{(0)}
\end{eqnarray}
which is real, and the second part
\begin{eqnarray}
\varepsilon_2=i\sum_j |\psi^R_j|^2 V_j\simeq i(\sum_{j} V_j)/\xi,
\end{eqnarray}
which obeys the Gaussian distribution by the central limit theorem. Here, the approximation  $|\psi^R_j|^2\simeq 1/\xi$ is used, which is better justified for small $q$, i.e., near the band center. In summary, the growing factor obeys the following distribution
\begin{eqnarray}
P(\lambda=\text{Im}\varepsilon_2)\propto e^{-\lambda^2/(2\sigma_\lambda^2)}
\end{eqnarray}
where 
\begin{eqnarray}
\sigma_\lambda\simeq\frac{\sigma_V}{\sqrt{\xi}}\sim\frac{W}{\sqrt{\xi}}.
\end{eqnarray}
Here, $\sigma_V=\sqrt{\langle V^2\rangle}=W/\sqrt{3}$. Note the analysis above gradually fails when $k$ deviates from $\pm \pi/2$ as the relation $|\psi^R_{j}|^2=1/\xi$ becomes invalid as it approaches the (real part of) band edge. In other words, the Gaussian distribution fits well for eigenvalues with its eigenstates having a long wavelength, which can meet the condition for the central limit theorem better. 

To further validate our prediction of an emergent Gaussian-type ImDOS, we numerically examine the distribution of the imaginary parts of eigenvalues under varying disorder strengths, as shown in Fig.~\ref{fig:GuassianEmergence}. We consider two types of disorder distributions: uniform and triangular. In both cases, we observe that as the disorder becomes sufficiently weak, the ImDOS indeed converges toward a Gaussian distribution.

\subsection{Intermediate disorder ($W\sim t_0$)}

\begin{figure}
    \centering
\includegraphics[width=0.4\linewidth]{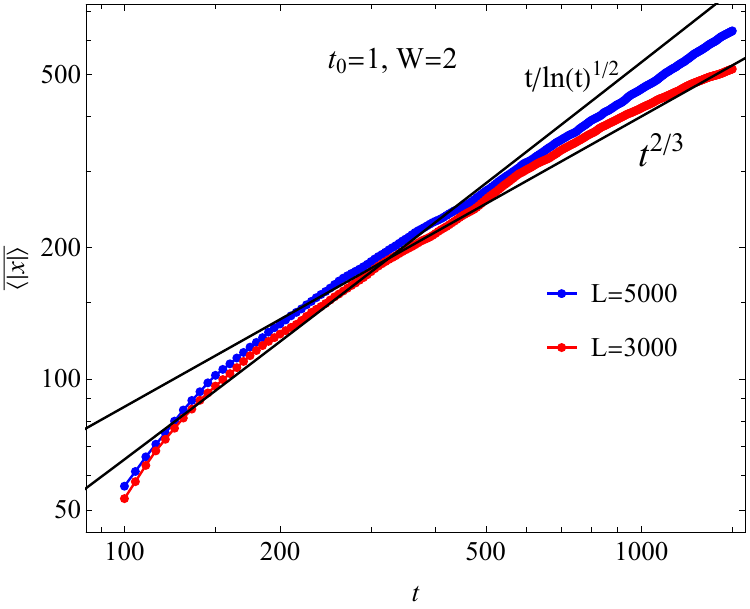}
    \caption{Scaling for intermediate disorder with $W=2t_0$, where $500$ realizations of disorder are performed.}
 \label{fig:crossover}
\end{figure}

In the presence of an intermediate disorder with $W\sim t_0$, the form of ImDOS can not be easily identified.  However, it is expected that the ImDOS would bridge the two extreme cases with strong or weak disorders, such that the long-time scaling in this scenario exhibits a crossover behavior. Indeed, via numerical simulation in Fig.~\ref{fig:crossover}, we find the scaling falls in between the scaling forms dominated by the Gaussian (weak-disorder limit) and the linear (strong-disorder limit) ImDOS.

\section{Renormalization group analysis}
In this section, we follow Ref.~\cite{Medina1989BurgereqRG} to carry out a dynamical renormalization group analysis. In the long-wavelength limit, the system is  governed by a Schr\"odinger equation 
\begin{eqnarray}\label{eq:SchrodingerEq}
 i\partial_t\psi(\mathbf x,t)=[-D\nabla^2+iV(\mathbf x)]\psi(\mathbf{x},t)   
\end{eqnarray}
where $D=\eta=t_0$ (set $\hbar=1$), and
\begin{eqnarray}\label{eq:disordercorrelation}
\langle V(\mathbf{x})V(\mathbf{x}^\prime)\rangle=v\delta^d(\mathbf x-\mathbf x^\prime)
\end{eqnarray}
with $v$ reflecting the disorder strength. 
Here, we should note that the correlator in Eq.~\eqref{eq:disordercorrelation} is valid for spatially uncorrelated disorder, regardless of the distribution function of the on-site potential variable. Specifically, all types of disordered potentials in the lattice model considered in the main text satisfy the following relations:
\begin{eqnarray}
&&\langle V(\mathbf x_{\{n_i\}})\rangle =0,\label{eq:disorderaverage}\\
&&\langle V(\mathbf x_{\{n_i\}})V(\mathbf x_{\{n_i^\prime\}})\rangle =v\prod_{i=1}^d\delta_{n_i,n_i^\prime}\label{eq:discretecorrlation},    
\end{eqnarray}
where $\mathbf x_{\{n_i\}}=\sum_{i=1}^d n_i\Delta x \mathbf e_i$ with $\mathbf e_i$, $\Delta x$ being the unit vector along $i$-direction and unit-cell length, respectively. Eq.~\eqref{eq:disorderaverage} is always satisfied, as the potential can be trivially shifted by any nonzero $\langle V(\mathbf x_{\{n_i\}})\rangle$ without affecting the localization properties. For $\mathbf x_{\{n_i\}}\neq \mathbf x_{\{n_i^\prime\}}$, uncorrelated disorder satisfies $\langle V(\mathbf x_{\{n_i\}})V(\mathbf x_{\{n_i^\prime\}})\rangle=\langle V(\mathbf x_{\{n_i\}})\rangle\langle V(\mathbf x_{\{n_i^\prime\}})\rangle=0$. When $\mathbf x_{\{n_i\}}=\mathbf x_{\{n_i^\prime\}}$, the correlator simplifies to $\langle V(\mathbf x_{\{n_i\}})V(\mathbf x_{\{n_i^\prime\}})\rangle=\langle [V(\mathbf x_{\{n_i\}}]^2)\rangle=\int dV V^2p(V)=v$, where $v$ depends on the disorder distribution function $p(V)$. In the long-wavelength limit, the unit-cell length $\Delta x$ can be considered infinitesimal. Therefore, the continuum version of Eq.~\eqref{eq:discretecorrlation} becomes Eq.~\eqref{eq:disorderaverage} by identifying $\mathbf x=\mathbf x_{\{n_i\}}$, $\mathbf x^\prime=\mathbf x_{\{n_i^\prime\}}$ and considering that
\begin{eqnarray}\label{eq:delta}
\lim_{\Delta x\rightarrow 0}\frac{1}{(\Delta x)^d}\prod_{i=1}^d\delta_{n_i,n_i^\prime}=\lim_{\Delta x\rightarrow 0}\frac{1}{(\Delta x)^d}\delta_{\mathbf x_{\{n_i\}},\mathbf x_{\{n_i^\prime\}}}=\delta^d(\mathbf x-\mathbf x^\prime). 
\end{eqnarray}
This relation follows from the fact that 
\begin{eqnarray}
g(\mathbf x_{\{n_i\}})=\sum_{\{n_i^\prime\}}g(\mathbf x_{\{n_i^\prime\}})\prod_{i=1}^d\delta_{n_i,n_i^\prime}\xrightarrow{\Delta x\rightarrow 0}\lim_{\Delta x\rightarrow 0}\int \frac{d^d\mathbf x^\prime}{(\Delta x)^d}\Big(\prod_{i=1}^d\delta_{n_i,n_i^\prime}\Big)g(\mathbf x^\prime)=\int d^d\mathbf x^\prime \delta^d(\mathbf x-\mathbf x^\prime) g(\mathbf x^\prime)=g(\mathbf x),
\end{eqnarray}
where $g(\cdots)$ is an arbitrary function of spatial coordinates.\\

Performing a Cole-Hopf transformation $\psi(\mathbf x,t)=\exp(\Phi(\mathbf x,t))$, the Schrodinger equation Eq.~\eqref{eq:SchrodingerEq} is converted to
\begin{eqnarray}\label{seq:nonlinear1}
\partial_t\Phi=iD\nabla^2\Phi+i\eta(\mathbf{\nabla}\Phi)^2+V(\mathbf{x}).
\end{eqnarray}
Turning to the momentum-frequency space via the Fouriour and Laplace transformation
\begin{eqnarray}\label{eq:FLtransformation}
   \Phi(\mathbf x,t)=\int_{\gamma-i\infty}^{\gamma+i\infty}\frac{ds}{2\pi i}\int_{k<\Lambda}\frac{d^dk}{(2\pi)^d} e^{i\mathbf k\cdot\mathbf x+st}\Phi(\mathbf{k},s)
\end{eqnarray}
where $\gamma\in\mathbbm{R}$ is chosen to make the integral kernel convergent, i.e., all singularities lying at the left of the line $z=\gamma$ on the complex plane,  
and
\begin{eqnarray}
V(\mathbf{k},s)=i2\pi\delta(s)V(\mathbf k),\qquad\text{with}
\quad
\langle V(\mathbf{k})V(\mathbf{k}^\prime)\rangle=v(2\pi)^d\delta^d(\mathbf k+\mathbf k^\prime).
\end{eqnarray} 
Here, the Laplace transformation is applied in consideration of the complex eigenvalues in our system. Substituting Eq.~\eqref{eq:FLtransformation} into
Eq.~\eqref{seq:nonlinear1} results in 
\begin{eqnarray}
\Phi(\mathbf{k},s)=G_0(\mathbf{k},s)V(\mathbf k,s)-i\eta G_0(\mathbf{k},s)\int_{\Omega,\mathbf q} \mathbf q\cdot(\mathbf k-\mathbf q)\Phi(\mathbf q,\Omega)\Phi(\mathbf k-\mathbf q,s-\Omega)
\end{eqnarray}
where $\int_{\Omega,\mathbf q}=\int_{\gamma-i\infty}^{\gamma+\infty}d\Omega/(i2\pi)\int d^dq/(2\pi)^d$ and
\begin{eqnarray}
G_0(\mathbf k,s)=\frac{1}{s+iDk^2}.
\end{eqnarray}
\begin{figure}
    \centering
    \includegraphics[width=0.8\linewidth]{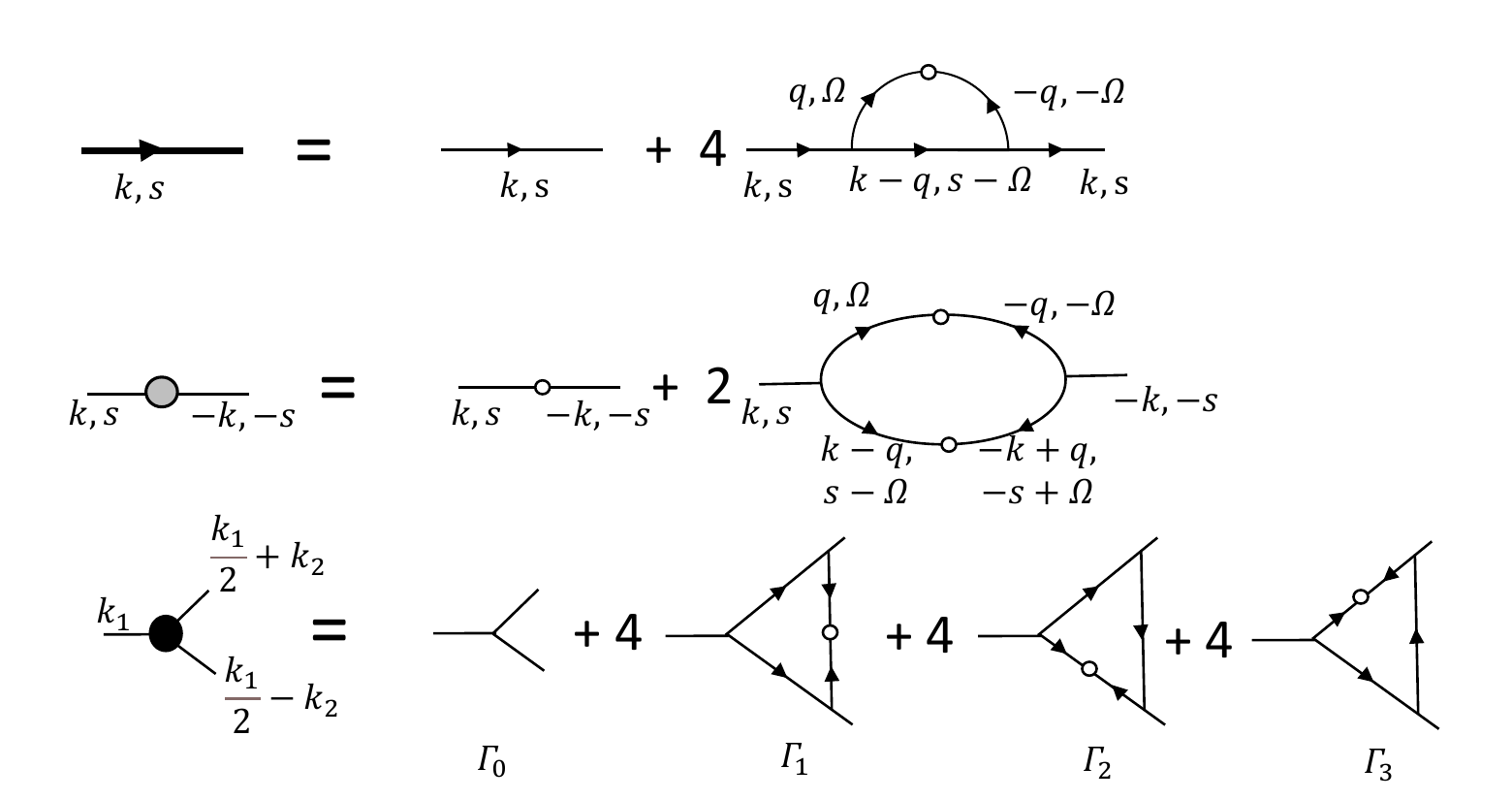}
    \caption{By averaging over disorders, the corresponding functions will be renormalized. In the diagrams above, corrections up to the third order of $\eta$ are considered.  The diagrams respectively describe the renormalized (Top) Green function, (Middle) correlator, and (Bottom) vertex. The thin and thick lines represent the bare and renormalized Green function, respectively; the thin line with a small open circle denotes the bare correlator, while the thin line with a gray solid disk stands for the renormalized correlator; the black solid disk is the renormalized vertex.}
    \label{fig:diagram_oneloop}
\end{figure}

The one-loop correction to the Green function is shown diagrammatically in Fig.~\ref{fig:diagram_oneloop}, which is written as
\begin{eqnarray}
   G(\mathbf k,s)=G_0(\mathbf k,s)+4(-i\eta)^2G_0(\mathbf k,s)\int_{\mathbf q,\Omega}\Big[(\frac{\mathbf{k}}{2}-\mathbf q)\cdot(\frac{\mathbf k}{2}+\mathbf q)\Big]\Big[\mathbf k\cdot(-\frac{\mathbf k}{2}+\mathbf q)\Big] G_0(\frac{\mathbf k}{2}+\mathbf{q},\frac{s}{2}+\Omega)C(\frac{k}{2}-\mathbf{q},\frac{s}{2}-\Omega),
\end{eqnarray}
and
\begin{eqnarray}
    C(\mathbf q,\Omega)=\int_{\omega,\mathbf{k}}\langle \Phi_0(\mathbf q,\Omega)\Phi_0(\mathbf k,\omega)\rangle=i2\pi vG_0(\mathbf q,\Omega)G_0(-\mathbf q,-\Omega)\delta(\Omega),
\end{eqnarray}
with $\Phi_0(\mathbf{q},\Omega)=G_0(\mathbf q,\Omega)V(\mathbf q,\Omega)$. Then second term in the limit $\mathbf{k}\rightarrow 0$ and $s\rightarrow 0$ is given as
\begin{eqnarray}
    I(\mathbf k,s)=&&\frac{-4v\eta^2G^2_0(\mathbf k,s)}{(2\pi)^d}\int_{\gamma-i\infty}^{\gamma+i\infty}d\Omega\int d^d q(\frac{k^2}{4}-q^2)(-\frac{k^2}{2}+kq\cos\theta)\frac{1}{\frac{s}{2}+\Omega+iD|\frac{\mathbf k}{2}+\mathbf q|^2}\nonumber\\
    &&\times \frac{1}{\frac{s}{2}-\Omega+iD|\frac{\mathbf k}{2}-\mathbf q|^2} \frac{1}{-\frac{s}{2}+\Omega+iD|\frac{\mathbf k}{2}+\mathbf q|^2}\delta(\frac{s}{2}-\Omega)\nonumber\\
    &&=\frac{-4v\eta^2}{(2\pi)^d}\frac{1}{(iDk^2)^2}\int d^d q\frac{(\frac{k^2}{4}-
    q^2)(-\frac{k^2}{2}+kq\cos\theta)}{(iD)^3(\frac{k^2}{2}+q^2+kq\cos\theta)(\frac{k^2}{2}+q^2+kq\cos\theta)^2}\nonumber\\
   &&=\frac{-4v\eta^2}{i(2\pi)^d}\frac{1}{D^5k^4}\int dq q^{d-1}S_{d-1}\int_0^\pi d\theta\sin^{(d-2)}\theta [q^{-4}k^2(\frac{1}{2}-\cos^2\theta)+\text{higer orders of $k$}]\nonumber\\
    &&=\frac{-v\eta^2K_d}{iD^5}\frac{2(d-2)}{d}\frac{1}{k^2}\int dq q^{d-5},
\end{eqnarray}
where we used $S_{d-1}\int_0^\pi d\theta\sin^{(d-2)}\theta\cos^2\theta=(S_{d-1}/d)\int_0^\pi \sin^{(d-2)}\theta=S_d/d$, and $K_d=S_d/(2\pi)^d$. By the results above, we obtain the renormalized parameter $\Tilde{D}$ is given by
\begin{eqnarray}
\Tilde{D}=D(1+\frac{2(d-2)v\eta^2K_d}{dD^4}\int dq q^{d-5}). 
\end{eqnarray}

The one-loop correction to correlator $C(\mathbf{k},s)$ is given by the second line in Fig.~\ref{fig:diagram_oneloop}, which further gives the correction to the disorder strength in the limit $\mathbf k\rightarrow 0$ and $s\rightarrow 0$ as below 
\begin{eqnarray}
\Tilde{v}&=&v+\lim_{\mathbf{k}\rightarrow 0, s\rightarrow 0}
\frac{2}{i(2\pi)}(-i\eta)^2\int_{\Omega,\mathbf{q}}[(\mathbf{k}-\mathbf{q})\cdot\mathbf q]^2C(\mathbf{q},\Omega)C(\mathbf k-\mathbf{q},s-\Omega)\nonumber\\
&=& v+2(-i\eta)^2 v^2(i2\pi)\frac{S_d}{i(2\pi)^{d+1}}\int dq q^{d-1}q^4 G_0(\mathbf q,0) G_0(-\mathbf q,0) G_0(-\mathbf q,0) G_0(\mathbf q,0)\nonumber\\
&=& v-\frac{2v^2\eta^2}{D^4}K_d\int dq q^{d-5}.
\end{eqnarray}
Here, the $i2\pi$ in the first line is put there to cancel extra $i2\pi$ in $C(\cdots)$ functions because it can be shown that to form one loop by averaging disorder, there only has one $i2\pi$ factor left even if higher order correlator, e.g. $\langle V VVV\rangle$, are evolved in, other factors cancel with vertex integral (generating $C(\mathbf q,\Omega)$) to close the loop. 

The third line gives corrections to $\eta$ as
\begin{eqnarray}
\Tilde{\eta}=\eta(1+\gamma_1+\gamma_2+\gamma_3),
\end{eqnarray}
with $\gamma_i$ corresponding the correction from $\Gamma_i$ ($i=1,2,3$). Specifically,
\begin{eqnarray}
\gamma_1&&=\lim_{\mathbf{k}_{1,2}\rightarrow 0}\lim_{s_{1,2}\rightarrow 0}\frac{\Gamma_1}{\Gamma_0}\nonumber\\
&&=\lim_{\mathbf{k}_{1,2}\rightarrow 0}\lim_{s_{1,2}\rightarrow 0}\frac{4(-i\eta)^3}{-i\eta(\frac{\mathbf k_1}{2}+\mathbf k_2)\cdot (\frac{\mathbf k_1}{2}-\mathbf k_2)}\int_{\Omega,\mathbf q}U_1(\mathbf k_1,\mathbf k_2,\mathbf q)G_0(\frac{\mathbf k_1}{2}+\mathbf q,\frac{s_1}{2}+\Omega)G_0(\frac{\mathbf k_1}{2}-\mathbf q,\frac{s_1}{2}-\Omega)C(\mathbf q-\mathbf k_1,\Omega-s_2)\nonumber\\
&&=4(-i\eta)^2v\int\frac{d^d q}{(2\pi)^d}\frac{1}{(iD q^2)^4}\frac{q^4(\frac{k_1^2}{2}\cos^2\theta_1-k_2^2\cos^2\theta_2)}{\frac{k_1^2}{4}-k_2^2}\nonumber\\
&&=-\frac{4\eta^2v}{dD^4}K_d\int dq q^{d-5}.
\end{eqnarray}
Here, we used 
\begin{eqnarray}
U_1(\mathbf k_1,\mathbf k_2,\mathbf q)&&=\Big[(\frac{\mathbf k_1}{2}+\mathbf q)\cdot(\frac{\mathbf k_1}{2}-\mathbf q)\Big]\Big[(\frac{\mathbf k_1}{2}-\mathbf k_2)\cdot(\mathbf k_2-\mathbf q)\Big]\Big[(\frac{\mathbf k_1}{2}+\mathbf k_2)\cdot(\mathbf q-\mathbf k_2)\Big]\nonumber\\
&&=-q^2(\mathbf k_2-\frac{\mathbf k_1}{2})\cdot\mathbf q (\mathbf k_2+\frac{\mathbf k_1}{2})\cdot\mathbf q+(\text{higher orders of $k_{1,2}$}).
\end{eqnarray}
Similarly, we can obtain the other two vertex corrections
\begin{eqnarray}
\gamma_3=\gamma_2&&=\lim_{\mathbf{k}_{1,2}\rightarrow 0}\lim_{s_{1,2}\rightarrow 0}\frac{\Gamma_2}{\Gamma_0}\nonumber\\
&&=\lim_{\mathbf{k}_{1,2}\rightarrow 0}\lim_{s_{1,2}\rightarrow 0}\frac{4(-i\eta)^3}{-i\eta(\frac{\mathbf k_1}{2}+\mathbf k_2)\cdot (\frac{\mathbf k_1}{2}-\mathbf k_2)}\int_{\Omega,\mathbf q}U_2(\mathbf k_1,\mathbf k_2,\mathbf q)G_0(\frac{\mathbf k_1}{2}+\mathbf q,\frac{s_1}{2}+\Omega)G_0(\mathbf q-\mathbf k_2)C(\frac{\mathbf k_1}{2}-\mathbf{q})\nonumber\\
&&=\frac{4\eta^2v}{dD^4}K_d\int dq q^{d-5}=-\gamma_1
\end{eqnarray}
where we omitted the frequency argument in the expression, and used
\begin{eqnarray}
U_2(\mathbf k_1,\mathbf k_2,\mathbf q)&&=\Big[(\frac{\mathbf k_1}{2}+\mathbf q)\cdot(\frac{\mathbf k_1}{2}-\mathbf q)\Big]\Big[(\frac{\mathbf k_1}{2}+\mathbf k_2)\cdot(\mathbf q-\mathbf k_2)\Big]\Big[(\mathbf q-\frac{\mathbf k_1}{2})(\frac{\mathbf k_1}{2}-\mathbf k_2)\Big]\nonumber\\
&&=q^2(\mathbf k_2-\frac{\mathbf k_1}{2})\cdot\mathbf q (\mathbf k_2+\frac{\mathbf k_1}{2})\cdot\mathbf q +(\text{higher orders of $k_{1,2}$}).
\end{eqnarray}
Summarizing three terms up, we obtain the dressed vertex is given by
\begin{eqnarray}
\Tilde{\eta}=\eta(1+\frac{4\eta^2v}{dD^4}K_d\int dq q^{d-5}).
\end{eqnarray}

It is obvious that all integrals induced by corrections are divergent for $d<4$. This difficulty can be resolved by performing a renormalization group (RG) analysis. 
Take the integral over a small shell $q\in[e^{-l},1]$ (setting $\Lambda=1$ for convenience) to obtain
\begin{eqnarray}
&&\Tilde{D}=D[1+\frac{2(d-2)v\eta^2K_d}{dD^4}\frac{1-e^{(4-d)l}}{(d-4)}],\nonumber\\
&&\Tilde{v}=v[1-\frac{2v\eta^2}{D^4}K_d\frac{1-e^{(4-d)l}}{(d-4)}],\nonumber\\
&&\Tilde{\eta}=\eta[1+\frac{4v\eta^2K_d}{dD^4}\frac{1-e^{(4-d)l}}{(d-4)}].
\end{eqnarray}

Now, we rescale the variables to compensate the integral over the momentum shell, $\mathbf x\rightarrow e^{l}\mathbf x$, $t\rightarrow e^{zl}t$, and $\Phi\rightarrow e^{\chi l}\Phi$. The modified dynamic equation becomes
\begin{eqnarray}
 \partial_t\Phi=ie^{(z-2)l}\Tilde{D}\nabla^2\Phi+ie^{(\chi+z-2)l}\Tilde{\eta}(\nabla\Phi)^2+e^{(z-\chi)l}\Tilde{V}(e^l\mathbf x).
\end{eqnarray}
By identifying the modified coefficient as a function of $l$ and taking $l=\delta l\ll 1$, we can obtain the following differential recursion relations
\begin{eqnarray}\label{eq:RGflow}
&&\frac{dD}{d l}=D(z-2+\frac{2(d-2)v\eta^2K_d}{dD^4}), \nonumber\\
&&\frac{dv}{dl}=v[2(z-\chi-\frac{d}{2})-\frac{2v\eta^2}{D^4}K_d],\nonumber\\
&&\frac{d\eta}{dl}=\eta(\chi+z-2+\frac{4v\eta^2K_d}{dD^4}).  
\end{eqnarray}
The RG equations above can yield the differential equation for a coupling constant $g=v\eta^2/D^4$,
\begin{eqnarray}
\frac{dg}{dl}=g[4-d+\frac{2(12-5d)}{d}K_dg].
\end{eqnarray}
There are two fixed points, $g_1^\ast=0$, $g^\ast_2=d(d-4)/[2(12-5d)K_d]$. It is easy to obtain that for $g_1=g_1^\ast+\delta g_1$
\begin{eqnarray}
    \frac{d\delta g_{1}}{dl}=(4-d)\delta g_{1}+\cdots,
\end{eqnarray}
which implies that $g_1^\ast$ is unstable for $d<4$.  For $d< 12/5$, $g^\ast_2<0$, which is unphysical because $g$ is positive by definition. However, for $d=3$, $g^\ast_2=1/(2K_d)$ is indeed a fixed point. At this fixed point, followed from Eq.~\eqref{eq:RGflow}, we obtain
\begin{eqnarray}
  z=5/3,\qquad \chi=2/3.  
\end{eqnarray}
Since $\chi<1$, this fixed point is stable against higher-order terms of $i\eta(\nabla\Phi)^2$ generated from RG process associated with Eq.~\eqref{seq:nonlinear1}.

In $d=3$, the long-time behavior for systems with weak disorder will be dictated by the new fixed point. Specifically, the displacement-time relation will be invariant under rescaling for a sufficiently long time and large spatial scales. Assuming $|\mathbf{x}|\sim t^\alpha$, the rescaling invariance requires that $|e^l\mathbf{x}|\sim (e^{zl}t)^\alpha$ is equivalent to the original scaling, leading to $\alpha=1/z$. Therefore, the $3d$ weak-disorder scaling is given by 
\begin{eqnarray}
 |\mathbf{x}|\sim t^{1/z}=t^{3/5}.   
\end{eqnarray}

\section{In the framework of the Lindblad master equation}
We consider general open systems with random losses described by the Lindblad master equation. It can be shown that the dynamics is described by the same scaling theory outlined in the main text.

\subsection{Lattice model with random dissipation}
We consider an open system described by the following master equation for the density matrix $\hat\rho$
\begin{eqnarray}\label{seq:master1}
\frac{d\hat \rho}{d t}=-i[\hat H_0,\hat \rho]+\sum_{\mathbf x}\gamma_{\mathbf x}\mathcal{D}[\hat a_{\mathbf x}]\hat \rho  
\end{eqnarray}
where $\hat H_0=\sum_{\mathbf x,\mathbf x^\prime}h_{\mathbf x,\mathbf{x}^\prime}\hat a^\dagger_{\mathbf x}\hat a_{\mathbf x^\prime}$ with $\mathbf{x},\mathbf x^\prime$ labeling lattice sites, and $\mathcal D[\hat O]\hat\rho=\hat O\hat\rho\hat O^\dagger-\frac{1}{2}(\hat O^\dagger\hat O\hat\rho+\hat\rho\hat O^\dagger\hat O)$ with $\hat O=\hat a_{\mathbf x}$.  On the r.h.s. of Eq.~\eqref{seq:master1}, the first term describes a unitary evolution given by a Hermitian Hamiltonian $\hat H_0$, and the second term contributes to local losses with site-dependent random rate $\gamma_{\mathbf x}$. For a time-independent operator $\hat A$, its averaged value satisfies the following equation
\begin{eqnarray}
   \frac{d\langle\hat A\rangle}{d t} &&=-i\text{Tr}([\hat H_0,\hat \rho]\hat A)+\sum_{\mathbf x}\gamma_{\mathbf x}\text{Tr}(\mathcal{D}[\hat a_{\mathbf x}]\hat \rho\hat A)\nonumber\\
   &&=-i\text{Tr}(\hat \rho[\hat A,\hat H_0])+\frac{1}{2}\sum_{\mathbf x} \gamma_{\mathbf x}\text{Tr}\Big(\hat \rho\hat a_{\mathbf x}^\dagger[\hat A,\hat a_{\mathbf x}]+\hat\rho[\hat a_{\mathbf x}^\dagger,\hat A]\hat a_{\mathbf x}\Big).
\end{eqnarray}

For our purpose, we consider a bilinear operator $\hat A=\hat a^\dagger_{\mathbf x_1}\hat a_{\mathbf x_2}$ whose average gives the two-point correlation function $G_{\mathbf x_1,\mathbf x_2}=\langle \hat a^\dagger_{\mathbf x_1}\hat a_{\mathbf x_2}\rangle=\text{Tr}[\hat\rho(\hat a^\dagger_{\mathbf x_1}\hat a_{\mathbf x_2})]$. Through a straightforward calculation, we obtain the dynamic equation of the correlation function 
\begin{eqnarray}
   \frac{d G_{\mathbf x_1,\mathbf x_2}}{d t}&&=i\sum_{\mathbf x^\prime}(h^T_{\mathbf x_1,\mathbf x^\prime}G_{\mathbf x^\prime,\mathbf x_2}-G_{\mathbf x_1,\mathbf{x}^\prime}h^T_{\mathbf{x}^\prime,\mathbf x_2}) -\frac{\gamma_{\mathbf x_1}+\gamma_{\mathbf x_2}}{2}G_{\mathbf x_1,\mathbf x_2}.
\end{eqnarray}
This equation can be reduced to the dynamical equation of correlation matrix $G=G_{\mathbf x_1,\mathbf x_2}|\mathbf x_1\rangle\langle \mathbf x_2|$ as below
\begin{eqnarray}
 \frac{dG}{dt}=i[(h^T+iV) G-G (h^T-iV)]=i( H G-G H^\dagger)
\end{eqnarray} 
where $H=h^T+iV$ describes a Hermitian system $h^T$ in a random imaginary disorder $iV$,  and $V$ is a diagonal matrix with $V_{\mathbf x,\mathbf{x}^\prime}=\gamma_{\mathbf{x}}\delta_{\mathbf x,\mathbf x^\prime}$. The general solution of the correlation matrix is 
\begin{eqnarray}
    G(t)=e^{iHt}G(0)e^{-iH^\dagger t}.
\end{eqnarray}
We are interested in the case that $N$ bosonic particles initially locate at at $\mathbf x=\boldsymbol{0}$, i.e., $G(0)=N|\boldsymbol{0}\rangle\langle\boldsymbol{0}|$, the correlation at time $t$ becomes
\begin{eqnarray}
G(t)=N|\psi(t)\rangle\langle\psi(t)|
\end{eqnarray}
with $|\psi(t)\rangle=e^{iHt}| \boldsymbol{0}\rangle$. Correspondingly, the particle spreading distance for survived particles in the system is evaluated as 
\begin{eqnarray}
\langle|\mathbf x|\rangle=\sum_{\mathbf x}|\mathbf x|G_{\mathbf x,\mathbf x^\prime}(t)/ \sum_{\mathbf x}G_{\mathbf x,\mathbf x^\prime}(t)=\sum_{\mathbf x}|\mathbf x||\langle\mathbf x|\psi(t)\rangle|^2/\sum_{\mathbf x}|\langle\mathbf x|\psi(t)\rangle|^2.
\end{eqnarray}
Since $|\psi(t)\rangle$ evolves by following the ``Hamiltonian" $H$, it can also be projected into the eigenbasis of $H$. Considering that $H$ describes a Hermitian Hamiltonian perturbed by an imaginary disorder, the calculation of the spreading distance is exactly mapped to the scenario we discussed before.  

\subsection{Purely dissipative lattice model}
It has been shown that the discrete jumpy motion also appears in a 1D purely dissipative lattice model in Ref.~\cite{Longhi2023LocLindblad}. Here, we generalize this model to higher dimensions and show that the corresponding dynamics can also be characterized by the scaling theory outlined in the main text. 

We consider a $d$-dimensional purely dissipative lattice model whose dynamics is governed by the master equation for the density matrix $\hat\rho$
\begin{eqnarray}\label{seq:master2}
    \frac{d\hat \rho}{d t}=\sum_{\mathbf x}\Gamma\mathcal D[\hat z_{\mathbf x}]\hat{\rho}+\gamma_{\mathbf x}\mathcal{D}[\hat a_{\mathbf x}]\hat \rho
\end{eqnarray}
where $\hat z_{\mathbf{x}}=\hat a_{\mathbf x}+\sum_{\boldsymbol{\delta}}\hat a_{\mathbf x+\boldsymbol{\delta}}$ with $\boldsymbol{\delta}$ being the unit vector in a $d$-dimensional lattice model. The two terms on the r.h.s. of Eq.~\eqref{seq:master2} describe dissipative couplings between adjacent sites with a rate $\Gamma$, and local losses with site-dependent random rate $\gamma_{\mathbf x}$, respectively. In this model, the unitary evolution part does not appear in the master equation due to the absence of a Hermitian Hamiltonian.
Similar as before,  the dynamics of  two-point correlation function $G_{\mathbf x_1,\mathbf x_2}=\langle \hat a^\dagger_{\mathbf x_1}\hat a_{\mathbf x_2}\rangle$ can be obtained by  a straightforward calculation 
\begin{eqnarray}
   \frac{d G_{\mathbf x_1,\mathbf x_2}}{d t}&&= -(2\Gamma+\frac{\gamma_{\mathbf x_1}+\gamma_{\mathbf x_1}}{2})G_{\mathbf x_1,\mathbf x_2}-\frac{\Gamma}{2}\sum_{\boldsymbol{\delta}}(G_{\mathbf x_1+\boldsymbol{\delta},\mathbf x_2}+G_{\mathbf x_1-\boldsymbol{\delta},\mathbf x_2}+G_{\mathbf x_1,\mathbf x_2+\boldsymbol{\delta}}+G_{\mathbf x_1,\mathbf x_2-\boldsymbol{\delta}})\nonumber\\
   &&-\frac{\Gamma}{2}\sum_{\boldsymbol{\delta}\neq\boldsymbol{\delta}^\prime}(G_{\mathbf x_1+\boldsymbol{\delta}-\boldsymbol{\delta}^\prime,\mathbf x_2}+G_{\mathbf x_1,\mathbf x_2+\boldsymbol{\delta}-\boldsymbol{\delta}^\prime}).
\end{eqnarray}
This equation can be packed into a matrix form as below
\begin{eqnarray}
 \frac{dG}{dt}=H G+ G H  
\end{eqnarray}
where 
\begin{equation}
    H=-\sum_{\mathbf{x}}(\Gamma+\frac{\gamma_{\mathbf{x}}}{2})|\mathbf x\rangle\langle \mathbf{x}|-\sum_{\mathbf x}\sum_{\boldsymbol{\delta}}\frac{\Gamma}{2}(|\mathbf x\rangle\langle\mathbf x+\boldsymbol{\delta}|+|\mathbf x\rangle\langle\mathbf x-\boldsymbol{\delta}|)-\sum_{\mathbf x}\sum_{\boldsymbol{\delta}\neq\boldsymbol{\delta}^\prime}\frac{\Gamma}{2}|\mathbf x\rangle\langle\mathbf x+\boldsymbol{\delta}-\boldsymbol{\delta}^\prime|.
\end{equation}
Here, $H$ describes a Hermitian Anderson model with strictly negative eigenvalues given that $\Gamma,\gamma_{\mathbf{x}}>0$. Directly solving the correlation function yields $G(t)=e^{H t}G(0)G^{Ht}$. Following the previous logic, the spreading distance is governed by a ``state" of form $|\psi(t)\rangle=e^{Ht}|\boldsymbol{0}\rangle$ for particles initially sitting at $\mathbf x=\boldsymbol{0}$. Since here the dissipative evolution (reflected as an imaginary time) is dictated by a Hermitian $H$, the corresponding scaling behavior will be dominated by the density of state (DOS) of $H$, instead of ImDOS. Otherwise, the analysis is analogous to the non-Hermitian case.

\bibliographystyle{apsrev}
\bibliography{disorder}